\documentclass[pre,twocolumn,a4paper]{revtex4}

\usepackage{graphicx}
\usepackage{subfigure}
\usepackage{url}

\graphicspath{{figs/}}

\newcommand{\ie}{i.e.}
 
\newcommand{\eg}{e.g.}

\newcommand{\ca}{Ca\(^{2+}\)}

\newcommand{\half}{\frac{1}{2}}
\newcommand{\expectation}[1]{\ensuremath{\left \langle #1 \right \rangle}}
\newcommand{\approaches}{\rightarrow}

\newcommand{\abs}[1]{\ensuremath{\left|#1\right|}}

\newcommand{\puncspace}{\enspace}
\newcommand{\mathcomma}{\puncspace ,}
\newcommand{\mathperiod}{\puncspace .}

\newcommand{\twoargs}[2]{\left(#1; #2\right)}

\newcommand{\temp}{\ensuremath{\alpha}}
\newcommand{\threshold}{\ensuremath{V_{0}}}
\newcommand{\numchannels}[1]{\ensuremath{Z_{#1}}}
\newcommand{\channelstate}{\ensuremath{Z}}
\newcommand{\nchannelstate}[1]{\ensuremath{\channelstate_{#1}}}
\newcommand{\expectednchannelstate}[1]{\expectation{\nchannelstate{#1}}}
\newcommand{\expectedstate}[3]{\ensuremath{
                        \expectednchannelstate{#1}\twoargs{#2}{#3}}}
\newcommand{\nchannelstatevar}[1]{\ensuremath{\variance{\nchannelstate{#1}}}}
\newcommand{\statevar}[3]{\ensuremath{\nchannelstatevar{#1}\twoargs{#2}{#3}}}
\newcommand{\responsewidth}{\ensuremath{W}}

\newcommand{\logistic}[1]{\ensuremath{\frac{1}{1+e^{-#1}}}}
\newcommand{\stdev}{\sigma}
\newcommand{\variance}[1]{\ensuremath{\stdev_{#1}^{2}}}

\newcommand{\openprob}[2]{\ensuremath{p\twoargs{#1}{#2}}}
\newcommand{\closeprob}[2]{\ensuremath{q\twoargs{#1}{#2}}}

\newcommand{\Vest}{\ensuremath{\hat{V}}}
\newcommand{\expectedVest}{\ensuremath{\langle\Vest\rangle}}
\newcommand{\decodedvalue}[1]{\ensuremath{\Vest_{#1}}}
\newcommand{\estimatedinput}[2]{\ensuremath{\expectedVest\twoargs{#1}{#2}}}
\newcommand{\expecteddiff}{\ensuremath{\varepsilon}}
\newcommand{\expecteddifference}[2]{\ensuremath{\expecteddiff\twoargs{#1}{#2}}}
\newcommand{\expectedsquaredifference}[2]
                            {\ensuremath{\expecteddiff^{2}\twoargs{#1}{#2}}}
\newcommand{\decodingvariance}[3]{\ensuremath
                    {\variance{\decodedvalue{#1}}\twoargs{#2}{#3}}}
\newcommand{\decodingerror}[1]{\ensuremath{\Delta\decodedvalue{#1}}}

\newcommand{\totalvariance}[3]{\ensuremath{\decodingerror{#1}^{2}\twoargs{#2}{#3}}}

\newcommand{\wraprefprepost}[3]{\wrapprepost{#1}{#2}{\ref{#3}}}
\newcommand{\formatrefplain}{\ref}
\newcommand{\formatrefparens}{\wraprefprepost{(}{)}}

\newcommand{\wrapprepost}[3]{{#1}{#3}{#2}}

\newcommand{\tagwithlabel}[2]{#1~#2}

\newcommand{\makelabeledcrossrefmacro}[4]
	{\newcommand{#3}{#1{#4}{#2}}}
\newcommand{\makecrossrefmaker}[3]
	{\newcommand{#1}{\makelabeledcrossrefmacro{#2}{#3}}}

\newcommand{\eqnrefformat}{\formatrefparens}
\newcommand{\eqnlabelbinding}{\tagwithlabel}
\newcommand{\eqnlabel}{eq.}
\newcommand{\Eqnlabel}{Eq.}
\newcommand{\eqnslabel}{eqs.}
\newcommand{\Eqnslabel}{Eqs.}

\newcommand{\eqnnum}{\eqnrefformat}
\makecrossrefmaker{\newlabeledeqnref}{\eqnlabelbinding}{\eqnnum}
\makecrossrefmaker{\newwordpluseqnref}{\tagwithlabel}{\eqnnum}

\newlabeledeqnref{\eqn}{\eqnlabel}
\newlabeledeqnref{\Eqn}{\Eqnlabel}
\newlabeledeqnref{\eqns}{\eqnslabel}
\newlabeledeqnref{\Eqns}{\Eqnslabel}

\newwordpluseqnref{\andeqn}{and}
\newwordpluseqnref{\througheqn}{through}

\newcommand{\figrefformat}{\formatrefplain}
\newcommand{\figlabelbinding}{\tagwithlabel}
\newcommand{\figlabel}{fig.}
\newcommand{\Figlabel}{Fig.}
\newcommand{\figslabel}{figs.}
\newcommand{\Figslabel}{Figs.}

\newcommand{\fignum}{\figrefformat}
\makecrossrefmaker{\newlabeledfigref}{\figlabelbinding}{\fignum}
\makecrossrefmaker{\newwordplusfigref}{\tagwithlabel}{\fignum}

\newlabeledfigref{\fig}{\figlabel}
\newlabeledfigref{\Fig}{\Figlabel}
\newlabeledfigref{\figs}{\figslabel}
\newlabeledfigref{\Figs}{\Figslabel}

\newwordplusfigref{\andfig}{and}
\newwordplusfigref{\throughfig}{through}

\newcommand{\sxnrefformat}{\formatrefplain}
\newcommand{\sxnlabelbinding}{\tagwithlabel}
\newcommand{\sxnlabel}{section}
\newcommand{\Sxnlabel}{Section}
\newcommand{\sxnslabel}{sections}
\newcommand{\Sxnslabel}{Sections}

\newcommand{\sxnnum}{\sxnrefformat}
\makecrossrefmaker{\newlabeledsxnref}{\sxnlabelbinding}{\sxnnum}
\makecrossrefmaker{\newwordplussxnref}{\tagwithlabel}{\sxnnum}

\newlabeledsxnref{\sxn}{\sxnlabel}
\newlabeledsxnref{\Sxn}{\Sxnlabel}
\newlabeledsxnref{\sxns}{\sxnslabel}
\newlabeledsxnref{\Sxns}{\Sxnslabel}
	
\newwordplussxnref{\andsxn}{and}
\newwordplussxnref{\throughsxn}{through}

\newcommand{\tblrefformat}{\formatrefplain}
\newcommand{\tbllabelbinding}{\tagwithlabel}
\newcommand{\tbllabel}{table}
\newcommand{\Tbllabel}{Table}
\newcommand{\tblslabel}{tables}
\newcommand{\Tblslabel}{Tables}

\newcommand{\tblnum}{\tblrefformat}
\makecrossrefmaker{\newlabeledtblref}{\tbllabelbinding}{\tblnum}
\makecrossrefmaker{\newwordplustblref}{\tagwithlabel}{\tblnum}

\newlabeledtblref{\tbl}{\tbllabel}
\newlabeledtblref{\Tbl}{\Tbllabel}
\newlabeledtblref{\tbls}{\tblslabel}
\newlabeledtblref{\Tbls}{\Tblslabel}
	
\newwordplustblref{\andtbl}{and}
\newwordplustblref{\throughtbl}{through}

\begin{document}

\title{Multiple Thresholds in a Model System of Noisy Ion Channels}
\date{\today}    

\newcommand{\affil}{\affiliation}
\author{Michael J. Barber}
\email{mjb@uma.pt}
\affil{Universidade da Madeira,
		Centro de Ci\^encias Matem\'aticas,
		Campus Universit\'ario da Penteada,
		9000-390 Funchal,
		Portugal}
\altaffiliation[Current address: ]{ARC systems research, 
				1 Donau-City-Stra{\ss}e, 
				1220 Vienna, 
				Austria}
\author{Manfred L. Ristig}
\email{ristig@thp.uni-koeln.de}
\affil{Institut f\"{u}r Theoretische Physik, 
	Universit\"{a}t zu K\"{o}ln,
	D-50937 K\"{o}ln,
	Germany}

\begin{abstract}
	Voltage-activated ion channels vary randomly between open and closed states, 
	influenced by the membrane potential and other factors. Signal transduction is 
	enhanced by noise in a simple ion channel model. The enhancement occurs in 
	a finite range of signals; the range can be extended using populations of 
	channels. The range increases more rapidly in multiple-threshold channel 
	populations than in single-threshold populations. The diversity of ion channels 
	may thus be present as a strategy to reduce the metabolic costs of handling a 
	broad class of electrochemical signals.
\end{abstract}


\maketitle


Voltage-activated ion channels  are essential elements in 
biological signal transduction, playing important roles in synaptic 
transmission, generation of neural action potentials, regulation of membrane potentials
and intracellular \ca{} concentrations, and other 
cellular functions \cite{DodFor:2004,ThoNel:2005,WhiRubKay:2000,Yel:2002}. 
The gating dynamics of the channels allow the 
nonconductive cell membrane to take part in electrical conduction 
and signaling. 
Channels vary between a conducting or open state 
and a nonconducting or closed state, with intermediary states in the transition 
being unstable and short-lived.  The transition between open and closed states
is influenced 
by a broad assortment of factors, principally the membrane potential, 
but also including hormones, toxins, protein kinases and phosphatases,  
and thermal fluctuations.    
Voltage-activated channels are functionally diverse in their sensitivity to depolarization; 
indeed, \textcite{LeeDauCriLacPerKloSchPer:1999} identify no fewer than five distinct
activation thresholds for 
\ca{} channels.

Channel gating dynamics are intrinsic stochastic 
transitions that depend strongly on external factors, so that the 
channel constitutes a single-molecule sensor or communication 
channel, transforming membrane potentials into noisy ionic 
currents. 
Noise can have surprising effects in many nonlinear systems.
Perhaps the best known of these is stochastic resonance (SR), wherein the presence of noise 
enhances the response of a thresholding system to a weak periodic signal 
(for a review, see Ref.~\cite{GamHanJunMar:1998}). SR has been experimentally 
demonstrated in a system of parallel ion channels \cite{BezVod:1995}, and studied theoretically in numerous systems (see, \eg, 
Refs.~\cite{MosPei:1995,BezVod:1997,GaiNeiColMos:1997,%
GoyHan:2000,JunShu:2001,SchGoyHan:2001,WenObe:2003}).
In this work, we use a simple ion channel model to further investigate the 
stochastic nature of voltage-activated ion channels, characterizing the limits 
that noise imposes on information transduction in individual channels and 
systems of channels.



We make use of 
a discrete model in which the channel switches between
distinct open and closed states, omitting the dynamics of the transition 
process. Such a discrete model captures the bistable nature of the 
channel dynamics. By omitting the transition dynamics, we assume that any
stimulus to the channel varies slowly compared to the time scale of channel
opening and closing. We can extend this assumption to a quasistatic 
approximation, where the channel is always in equilibrium, and describe 
the channel opening probability by using the steady state 
(time \(t \approaches \infty\)) probability for the permissive
state \cite{JohWu:1995}, also called the activation function of 
the channel. This probability is given by
\begin{equation}
    P_{\infty} = \logistic{zF(V-\threshold)/RT}
    \mathcomma
    \label{eq:steadystateprob}
\end{equation}
where \( z \) is the (dimensionless) valence of the ``gating particles,'' 
\( F\) is Faraday's constant, \( V \) is the transmembrane potential, \( R \) 
is the ideal gas constant, and \( T \) is the temperature. 
The parameter \(\threshold\)
is a bias (or noisy threshold) in the potential to which the 
channel tends to open. 
\textcite{GaiNeiColMos:1997} have  investigated noise in model ion channels 
essentially identical to the ones defined by \eqn{eq:steadystateprob}, demonstrating 
that stochastic resonance occurs. 

For notational clarity, we lump several of the parameters into a thermal noise
parameter \temp, such that
\begin{equation}
    \temp = \frac{RT}{zF}
    \mathperiod
    \label{eq:tempabbrev}
\end{equation}
The definition in \eqn{eq:tempabbrev} amounts to changing the units of
measure for the temperature to Volts. At room temperature and with \(z = 1\), 
\temp\ is approximately 25 mV. 

Using \eqn{eq:tempabbrev}, the channel opening probability becomes
\begin{equation}
    \openprob{V}{\temp} = \logistic{(V - \threshold) / \temp}
    \mathcomma
    \label{eq:openprob}
\end{equation}
with a corresponding probability of remaining closed of \(\closeprob{V}{\temp} 
= 1 - \openprob{V}{\temp}\).
For single channels, or populations of channels with 
homogeneous behavior, we can take \(\threshold=0\) without loss of generality;
the behavior for other values of \(\threshold\) can be recovered by translating
the potential in \eqn{eq:openprob} by the desired value for the
threshold potential. In heterogeneous populations of channels, the differing 
values of the threshold potentials can have a profound effect on the behavior
of the system of channels, as we will demonstrate below.

For a system of \(N\) channels, each of 
which are exposed to the same transmembrane potential \(V\) but with different 
realizations of the noise (\ie, we have i.i.d.~noise), we let 
\numchannels{N} be the number of channels that open during the time interval.
The expected state of the membrane, \ie, \numchannels{N}, can 
be calculated in a straightforward fashion. The expectation 
value \(\expectedstate{N}{V}{\temp}\) and 
variance \(\statevar{N}{V}{\temp}\) can be expressed
using \openprob{V}{\temp}, \closeprob{V}{\temp}, and \(N\), 
giving
\begin{eqnarray}
    \expectedstate{N}{V}{\temp} & = & N \openprob{V}{\temp} \label{eq:expectedstateN} \\
    \statevar{N}{V}{\temp} & = & N \openprob{V}{\temp} \closeprob{V}{\temp}
    \label{eq:statevarianceN}
    \mathperiod
\end{eqnarray}
In calculating \eqns{eq:expectedstateN} \andeqn{eq:statevarianceN}, we have made use of
the independence of the noise for the channels.


We will focus on the behavior of ion channels near the threshold value. 
To explore the ability of the ion 
channels to serve as a transducer of electrical signals, we reproduce the input potential 
by decoding the state of
the membrane (\ie, the numbers of open and closed channels). 
Near the threshold, this gives rise to linear decoding rules. 
The basic approach is similar to the  ``reverse reconstruction''  using linear filtering 
that has been applied with great effect to the analysis of a number of biological systems
(see, \eg, Refs.~\cite{BiaRie:1992,BiaRieRuyWar:1991,RieWarRuyBia:1997,%
PraGabBra:2000,TheRodStuClaMil:1996}).   Despite the quite different nature 
of the signals we consider, our static population analysis (below) 
shares the key features of the temporal analyses in the
cited works. Specifically, by comparing the actual input to an 
estimate derived from the systemic response, we can work at greater remove
from the details of the systems.
We thus bypass the need for complete
descriptions of the transmission process and of the manner in which the 
system response is used. This permits general limits to be established on the 
signal-transducing ability of a population of channels, regardless of what 
the actual mechanisms may be.  

\begin{figure}
    \centering
    \includegraphics[width=8cm]{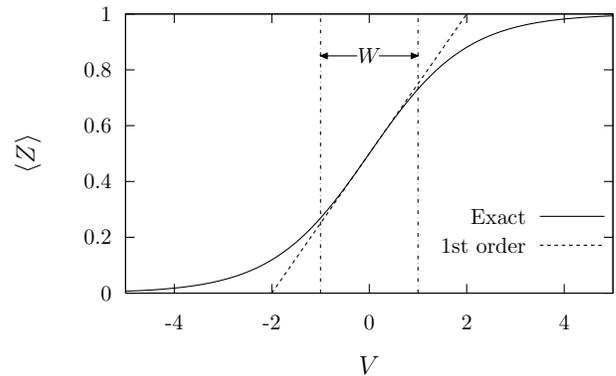}
    \caption[First order approximation of the channel opening expectation.]
    {First order approximation of the expectation value for the channel opening. 
    Near the threshold potential ($\threshold = 0$) of the channel, the expectation 
    is nearly linear. Further from the threshold, the expectation value saturates 
    at either zero or one, where the linear approximation diverges from the true 
    value. The values shown here are based on thermal noise $\temp=1$, giving
    a response width $\responsewidth=2$.}
    \label{fig:firstorderexpectationV}
\end{figure}

We 
expand the expected number of open channels \expectedstate{N}{V}{\temp} to first order 
near the threshold (\ie, $V \approaches 0$), giving 
\begin{equation}
    \expectedstate{N}{V}{\temp} = 
            \frac{N}{2} + \frac{N}{4\temp}V + O\left(V^{2}\right)
    \mathperiod
    \label{eq:stateexpansion}
\end{equation}
An example of the linear approximation is shown in 
\fig{fig:firstorderexpectationV}.

Dropping the higher order terms and inverting \eqn{eq:stateexpansion} 
suggests a linear decoding rule of the form
\begin{equation}
    \decodedvalue{N} = 4\temp\left(\frac{\numchannels{N}}{N} - \half \right) 
    \mathcomma
    \label{eq:decodingrule}
\end{equation}
where \(\decodedvalue{N}\) is the estimate of the input potential. Combining 
\eqns{eq:expectedstateN} \andeqn{eq:decodingrule}, we can show that
\begin{equation}
    \expectation{\decodedvalue{N}}(V; \temp) = 
                4\temp\left(\openprob{V}{\temp} - \half \right) 
    \mathperiod
    \label{eq:meandecodedsignal}
\end{equation}
The expected value of 
\(\decodedvalue{N}\) is thus seen to be independent of \(N\); for notational 
simplicity, we drop
the subscript and write \expectedVest.
Note that, as the thermal noise
increases, the 
expected value of the decoded potential closely matches the input potential over 
a broader range.

We must also consider the uncertainty of the potential value decoded from the state
of the membrane. This leads to a total decoding error \decodingerror{N} with the form
\begin{eqnarray}
    \totalvariance{N}{V}{\temp} & = & 
    			\expectation{\left( \decodedvalue{N} - V \right)^{2}} \nonumber \\
            & = & \expectedsquaredifference{V}{\temp} + \decodingvariance{N}{V}{\temp} 
    \mathcomma
    \label{eq:decodingerror}
\end{eqnarray}
where
\begin{eqnarray}
    \expecteddiff(V; \temp) & = & \expectedVest(V; \temp) - V
    \label{eq:expecteddifference} \\
    \decodingvariance{N}{V}{\temp} & = & 
        \expectation{\left( \decodedvalue{N} 
                - \expectedVest(V ; \temp) \right) ^{2} } \nonumber \\
    & = & \frac{16\temp^{2}}{N}
                \openprob{V}{\temp}\closeprob{V}{\temp} 
    \mathperiod
    \label{eq:decodingvariance}    
\end{eqnarray}

\begin{figure}
    \centering
    \includegraphics[width=8cm]{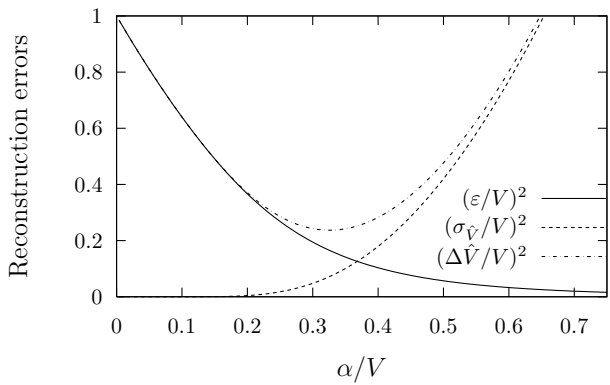}
    \caption[Comparison of decoding error sources.]
    {Comparison of decoding error sources. The values shown here are calculated
    for a single channel. The minimum in $\decodingerror{}^{2}$ occurs for 
    nonzero thermal noise, analogous to stochastic resonance.}
    \label{fig:decodingerrorcompare}
\end{figure}

Using \eqn{eq:openprob}, we can derive  several properties that are useful
for understanding the role of noise in the channel behavior. In particular, it 
is straightforward to show that 
\(\estimatedinput{V}{\temp} = V\estimatedinput{1}{(\temp/V)}\),
\(\expecteddifference{V}{\temp} = V\expecteddifference{1}{(\temp/V)}\),
and 
\(\decodingvariance{N}{V}{\temp} = V^{2} \decodingvariance{N}{1}{\temp/V}\), 
for all \(V \neq 0\).
Thus, the noise dependence for both the reconstructed stimulus and its 
uncertainty can be 
understood with a single stimulus. 
The total error \(\totalvariance{N}{1}{(\temp/V)}\) is minimized for a 
nonzero value of the noise parameter \(\temp\), analogous to the stochastic 
resonance effect; see \fig{fig:decodingerrorcompare}.

\begin{figure}
    \centering
    \includegraphics[width=8cm]{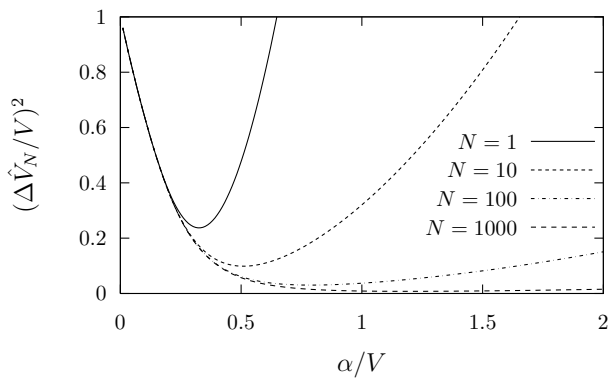}
    \caption[Effect of the number of channels on the decoding error.]
    {Effect of the number of channels on the decoding error. As $N$ becomes 
    large, the error curve flattens out, indicating a broad range of noise 
    values that all give similar accuracy in the decoding process.}
    \label{fig:decodingerrorN}
\end{figure}

In \fig{fig:decodingerrorN}, we show how 
\(\decodingerror{N}^{2}\) varies with the number of channels \(N\). 
As \(N\) increases, the error curve 
flattens out into a 
broad range of similar values. Thus, the presence of noise enhances signal 
transduction without requiring a precise relation between \(V\) and \temp.
This effect is analogous to the ``stochastic 
resonance without tuning'' first reported by \textcite{ColChoImh:1995} and 
previously explored in two-state systems not dissimilar to the one investigated here
(see, \eg, Ref.~\cite{NeiSchMos:1997}).

Informally stated, SR without tuning allows for a wider range of potentials 
to be accurately decoded from the channel states for any particular value of 
\temp. To make this notion of ``wider range'' precise, we again focus our 
attention on the expected behavior of the channels (see 
\fig{fig:firstorderexpectationV}). The expected channel response 
\expectation{\channelstate} 
matches well with the linear approximation when \(\abs{V} < \temp\). From this, 
the width \(\responsewidth\) can be defined 
to be \(2 \temp\).
Other definitions for the response width 
are, of course, possible, but we still should observe that the width is 
proportional to \(\temp\), since the probability for 
channel opening depends only on the ratio of \(V\) and \(\temp\) (\eqn{eq:openprob}). 
The same width is 
found for multiple identical channels, because the total expected current is 
proportional to the single channel behavior, without broadening the curve in 
\fig{fig:firstorderexpectationV}. 

The response width can thus be increased by increasing the thermal noise 
parameter~\temp. As seen in 
\figs{fig:decodingerrorcompare} \andfig{fig:decodingerrorN}, such an increase 
ultimately leads to a growth in the decoding error 
\(\decodingerror{N}^{2}\). As \temp\ becomes large, 
\(\decodingerror{N}^{2}\) is dominated by \variance{\decodedvalue{N}} and
we have the asymptotic behavior 
\begin{equation}
    \totalvariance{N}{V}{\temp} = O\left( \frac{\temp^{2}}{N} \right) 
    \mathcomma
    \label{eq:asymptoticerror}
\end{equation}
based on \eqn{eq:decodingvariance}.
The growth in \(\decodingerror{N}^{2}\) with increasing \temp\ thus can be 
overcome by further increasing the number of channels in the population. 
Therefore, the response width \(\responsewidth\) is effectively constrained by 
the number of channels \(N\), with \(W=O(\sqrt{N})\) for large \(N\). 

An arbitrary response width can be produced by assembling enough channels. 
However, this approach is inefficient, and greater width increases can be 
achieved with the same number of channels. Consider instead dividing up the total 
width into \(M\) subranges. These subranges can each be covered by 
a subpopulation of \(N\) channels, with the subpopulations having different thresholds from
one another. The width of each subrange is \(O(\sqrt{N})\), 
but the total width is \(O(M\sqrt{N})\). Thus, the total response width can increase 
more rapidly as additional types of channels are added.
Conceptually, multiple types of channels 
arise naturally as a way to provide a wide range of accurate responses, with 
multiple channels in each type providing independence from any need to ``tune'' 
the thermal noise to a particular level.

To describe the behavior of channels with different thresholds, much 
of the preceding analysis can be directly applied  by translating the functions 
along the potential axis to obtain the desired threshold. 
However, 
system behavior was previously explored near the threshold
value, but  heterogeneous populations of channels have multiple thresholds. Nonetheless, 
we can produce a comparable system by simply assessing system behavior
near the center of the total response width.

To enable a clean comparison, 
we set the thresholds in the heterogeneous populations so that a linear decoding rule can be produced. A simple approach that achieves this is to space the thresholds of the subpopulations 
by \(2\responsewidth = 4\temp\), with all channels being otherwise equal. The subpopulations with lower thresholds provide an upward shift in the expected number of open channels for higher threshold subpopulations, such that the different subpopulations are all approximated to first order by the same line. Thus, the expected total number of open channels leads to a 
linear decoding rule by expanding to first order and inverting, as was done 
earlier
for homogeneous populations. Note that this construction requires 
no additional assumptions about how the channel states are to be interpreted. 

To illustrate the effect of multiple types of channels, we begin with a 
homogeneous baseline population \(M=1\) of \(N=1000\) channels with 
\(\threshold=0\) and apply a potential \(V\) with thermal noise \(\temp=1\). 
Using the definition above, the response width is \(\responsewidth=4\). We then 
consider two 
cases, homogeneous and heterogeneous, in each of which we increase the response 
width by doubling the number of channels while maintaining similar error 
expectations for the decoded currents. 

\begin{figure}
    \centering
    \includegraphics[width=8cm]{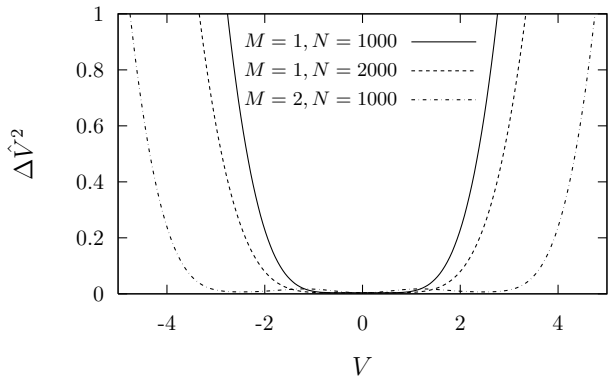}
    \caption[Total decoding error in homogeneous and heterogeneous
    populations of channels.]
    {Total decoding error in homogeneous and heterogeneous populations of 
    channels. The heterogeneous channel population ($M=2, N=1000$) has a 
    broader ``basin'' of low error values than the baseline ($M=1, N=1000$) and 
    homogeneous ($M=1, N=2000$) populations.}
    \label{fig:heterogeneousdecodingerror}
\end{figure}

In the homogeneous case, we have a single population (\(M=1\)) with \(N=2000\) 
channels. Doubling the number of channels allows us to 
increase the temperature parameter \(\alpha\) by a factor of \(\sqrt{2}\)
with similar expected errors outside the response width. Thus, we 
observe an extended range, relative to the baseline case, in which we can 
reconstruct the stimulus potential from the state of the channels 
(\fig{fig:heterogeneousdecodingerror}).

In the heterogeneous case, we instead construct two subpopulations (\(M=2\)) with 
\(N=1000\) channels. We leave the thermal noise parameter unchanged at \(\temp=1\). One of 
the subpopulations is modified so that the channel thresholds lie at 
\(+\responsewidth=2\), 
while the other is modified so that the channel thresholds lie at 
\(-\responsewidth=-2\). 
The resulting system of channels has a broad range in which we can 
reconstruct the stimulus potential with low error, markedly superior to the 
baseline and homogeneous cases 
(\fig{fig:heterogeneousdecodingerror}).
The approach used in this example
can be directly extended to three or more subpopulations.

The foregoing analysis and example 
suggest that the diversity of channel types found in a living 
cell are present as an information processing strategy, providing a 
means to efficiently handle a broad class of electrochemical signals.
The superior scaling
properties of heterogeneous populations of channels can have a profound impact on the 
cellular metabolism; large numbers of channels imply a large energetic investment, both in terms
of  the proteins needed to construct the channels and of the increased demand on ion pumps 
that accompanies the greater ionic currents \cite{WhiRubKay:2000}. The action potentials 
generated in neurons can require a significant energetic cost \cite{LauRuyAnd:1998}, making the 
tradeoff between reliably coding information and the metabolic costs potentially quite important.

In this picture, we expect that different types of cells will 
require different numbers of functionally different ion channels.  
Cells that perform sophisticated signaling and respond to a broad variety of 
signals will need a large number of functionally different ion channels, while cells that are 
more specialized to a narrower class of signals are likely to have a 
smaller number of functional types.  This appears to be generally consistent 
with the comparatively large variety of ion channels found in excitable cells
such as neurons \cite{Yel:2002}.  

It is interesting to reconsider the work of \textcite{LeeDauCriLacPerKloSchPer:1999} in light of the 
analysis presented here.  In their work, five \ca{} channels with distinct thresholds are tabulated, 
including three T-type channels interpreted as being important for the electrical responsiveness 
of neurons. However, a full understanding of the physiological roles of the channels remains to be 
found. The present analysis indicates that signal transduction by any given type of channel is intrinsically limited, regardless of the details of how the state of the channels is used. The  multiple types of
channels with their various thresholds provide a metabolically and evolutionarily favorable 
means to overcome those limits.  

Although we have used a specific model consisting of channels with thermal 
fluctuations modulating an input potential, we expect that the key result is 
more widely applicable. The demonstration of the advantage of multiple channel 
types largely follows from two factors that are not specific to the model 
channels. First, the distance of the input potential from the threshold is proportional 
to the level of the thermal noise, and, second, the total variance of the inputs to the 
channels is proportional to the number of channels. Ultimately, a multiplicity of 
functional types of channels with varying thresholds arises because the independently 
distributed noise provides 
a natural scale for the system. 

\begin{acknowledgments}
We would like to acknowledge support from the Portuguese Funda\c{c}\~ao para a Ci\^encia e 
a Tecnologia under Bolsa de Investiga\c{c}\~ao SFRH/BPD/9417/2002 and 
Plurianual CCM.

\end{acknowledgments}


\bibliographystyle{apsrev}

%

\end{document}